\documentclass[fleqn,usenatbib]{mnras}

\usepackage{newtxtext,newtxmath}

\usepackage[T1]{fontenc}
\usepackage{centernot}
\DeclareRobustCommand{\VAN}[3]{#2}
\let\VANthebibliography\thebibliography
\def\thebibliography{\DeclareRobustCommand{\VAN}[3]{##3}\VANthebibliography}

\usepackage{graphicx}	
\usepackage{amsmath}	
\usepackage{subcaption}

\title[Simulating Cold Dark Matter with PIKANs]{Solving the Cosmological Vlasov–Poisson Equations
with Physics-Informed Kolmogorov-Arnold Networks}

\author[N. Cerardi et al.]{
Nicolas Cerardi$^{1}$\thanks{E-mail: nicolas.cerardi@epfl.ch},  Emma Tolley$^{1}$, Ashutosh Mishra$^{1}$
\\
$^{1}$Institute of Physics, Laboratory of Astrophysics, École Polytechnique Fédérale de Lausanne (EPFL), Switzerland
}

\date{Accepted 2025 December 11. Received 2025 December 11; in original form 2025 October 2}

\pubyear{\the\year{}}

\begin{document}
\label{firstpage}
\pagerange{\pageref{firstpage}--\pageref{lastpage}}
\maketitle

\begin{abstract}

Cold dark matter (CDM) evolves as a collisionless fluid under the Vlasov–Poisson equations, but N-body simulations approximate this evolution by discretising the distribution function into particles, introducing discreteness effects at small scales. We present a physics-informed neural network approach that evolves CDM fields without any use of N-body data or methods, using a Kolmogorov–Arnold network (KAN) to model the continuous displacement field for one-dimensional halo collapse. Physical constraints derived from the Vlasov–Poisson equations are embedded directly into the loss function, enabling accurate evolution beyond the first shell crossing. The trained model achieves sub-percent errors on the residuals even after seven shell crossings and matches N-body results while providing a resolution-free representation of the displacement field. In addition, displacement errors do not grow over time, a very interesting feature with respect to N-body methods.
It can also reconstruct initial conditions through backward evolution when sufficient final-state information is available. Although current runtimes exceed those of N-body methods, this framework offers a new route to high-fidelity CDM evolution without particle discretisation, with prospects for extension to higher dimensions.

\end{abstract}

\begin{keywords}
gravitation -- dark matter -- simulations
\end{keywords}



\section{Introduction}
\label{introduction}

Dark matter (DM) constitutes 85\% of the matter content of the Universe~\citep{planck_collaboration_planck_2020} and drives the evolution of the cosmic large-scale structure, but its nature remains almost completely unknown.

\begin{figure}
	\centering 
	\includegraphics[width=0.49\textwidth]{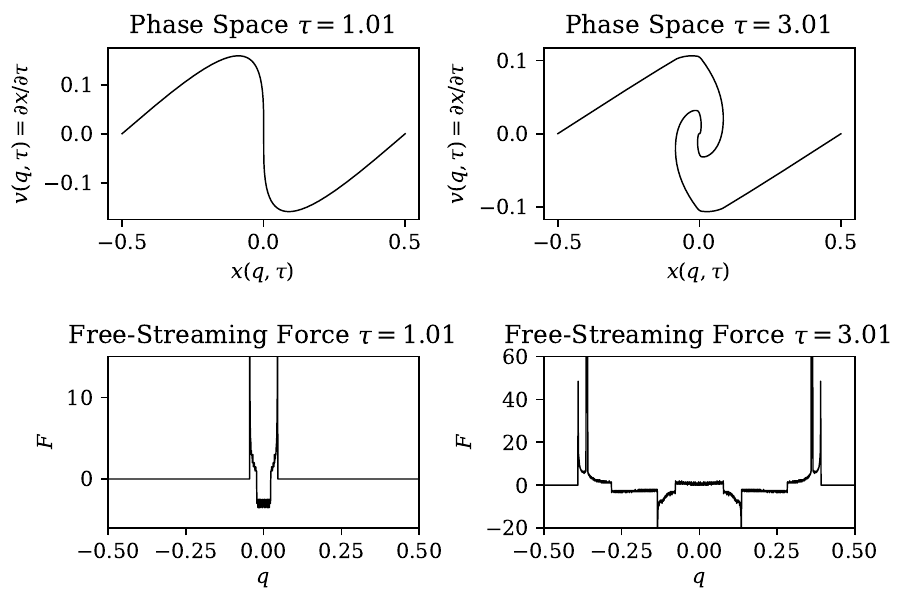}	
	\caption{The 2D phase space and the multi-stream force of a 1D N-body simulation with $N=4096$ particles at $a = \tau = 1.01$ (shortly after the first shell-crossing) and $a = \tau = 3.01$. The vanishing velocity dispersion of DM confines its evolution to a single line within 2D phase space. The singular density distribution created by the shell-crossing results in a discontinuous multi-stream force. The N-body simulation code is from \protect\cite{rampf_unveiling_2021}.} 
 \label{fig:1dnbodysim}%
\end{figure}

Aside from gravitational interactions, DM is extremely weakly interacting~\citep{markevitch_direct_2004, rocha_cosmological_2013} and is considered collisionless on cosmological scales. Let us define comoving coordinates $\vec x$, peculiar velocity $\vec v$, and conformal time $\tau$ related to the cosmic time $t$ via the cosmic scale factor $a$ such that ${\rm d}t = a(\tau) {\rm d} \tau $. For simplicity, we conduct our study under a flat and dark matter-dominated Universe, also known as Einstein-de Sitter.
The DM distribution $f = f(\vec x, \vec v, \tau)$ evolves according to the cosmic Vlasov–Poisson equations~\citep{bernardeau_large-scale_2002}:
\begin{equation}\label{eq:vp}
    \frac{\text{d}f}{\text{d} \tau} = 0 = \frac{\partial f}{\partial \tau} + \vec v \cdot \vec \nabla_x f
    - \vec \nabla_x \Phi \cdot \frac{\partial f}{\partial \vec v} ~,
\end{equation}
where $\Phi$ is the cosmological gravitational potential, sourced only by density fluctuations,
\begin{equation}
\nabla_x^2 \Phi = 4 \pi G a^2 \bar \rho(\tau) \delta(\vec x,\tau)~,
\end{equation}
 and $\delta(\vec x,\tau) $ is the matter density contrast defined from the matter density field $\rho$:
\begin{equation}
\rho = \bar \rho(\tau) (1 + \delta(\vec x,\tau)) = \int \text{d}v^3 f(\vec x, \vec v, \tau) ~,
\end{equation}
Equation~(\ref{eq:vp}) is very difficult to solve, as it is a non-linear partial differential equation involving seven variables.
Furthermore, cosmological observations indicate that the initial dark-matter distribution is extremely cold, which confines DM to a three-dimensional hypersurface within the full 6D phase-space, called the Lagrangian sub-manifold~\citep{rampf_cosmological_2021}, shown for 1+1D phase space in Figure~\ref{fig:1dnbodysim}.

Instead of solving for the dynamics of DM across the full 6D phase space, we can describe the dynamics of the manifold with a 3D displacement field $\vec \zeta(\vec q,\tau)$ which maps initial $(\tau = 0)$ particle positions $\vec q$ into the Eulerian particle  positions $\vec x(\vec q,\tau)$:
\begin{equation}\label{eq:Lmap}
\vec q ~ \mapsto ~ \vec x( \vec q,\tau) = \vec q + \vec \zeta(\vec q,\tau)~.
\end{equation}
The equation of motion for the particle trajectories $\vec x(q,\tau)$ is then~\citep[for a proper derivation see][]{uhlemann_semiclassical_2019}:
\begin{equation} \label{eq:vpL}
\frac{\partial^2 \vec x}{\partial \tau^2} + \frac{3}{2\tau} \frac{\partial \vec x}{\partial \tau} =  - \frac{3}{2 \tau} \vec \nabla_x \phi~,
\end{equation}
where $\vec \nabla_x$ the gradient operator in
Eulerian coordinates $\vec x$ and $\phi$ is the rescaled gravitational potential:
\begin{align}
    &\nabla_x^2 \phi = \frac{\delta(\vec x( \vec q,\tau))}{\tau},\label{eq:vpL1}\\
    &\delta(\vec x( \vec q,\tau)) = \hspace{-0.2em}\int \hspace{-0.5em} \text{d}q'^3 \delta_\text{D}[\vec x(\vec q, \tau) - \vec x(\vec q', \tau)] - 1, \label{eq:vpL2}
\end{align}
where $\delta_\text{D}$ is the Dirac delta.
When the dark-matter sheet begins to fold, called {\emph shell-crossing}, i.e. when fluid elements with different initial positions $\vec q$ end up at the same Eulerian position $\vec x$ through the mapping in Equation~(\ref{eq:Lmap}), the density field becomes singular and the description of dynamics in terms of an injective mapping no longer holds, as $\vec x(\vec q_1) = \vec x(\vec q _2) \centernot\implies \vec q_1 = \vec q _2$~\citep{rampf_unveiling_2021}.

The dynamics of Equations~(\ref{eq:vpL},\ref{eq:vpL1},\ref{eq:vpL2}) are often solved numerically using N-body methods~\citep{bertschinger_simulations_1998,colombi_dynamics_2001} such as particle mesh ~\citep[PM;][]{doroshkevich_two-dimensional_1980}, treecodes~\citep{appel_efficient_1985}, and adaptive mesh refinement~\citep[AMR;][]{villumsen_new_1989}. All N-body methods represent the DM phase-space distribution function as an ensemble of particles, i.e. a set of Dirac functions in phase-space interacting through gravitational forces, which can have non-trivial consequences on the numerical behaviour of the system~\citep{hernquist_are_1990,goodman_exponential_1993,splinter_fundamental_1997, worrakitpoonpon_violent_2024}.

For these reasons, it is important to confirm results obtained with traditional N-body methods with alternate numerical routes for solving the Vlasov–Poisson equations. For the case of cold DM (CDM), one strategy is to follow the evolution of Lagrangian phase-space elements to capture the foldings of the CDM fluid sheet~\citep{abel_tracing_2012, shandarin_cosmic_2012}. \cite{hahn_new_2013} incorporated this method into a Particle-Mesh approach to solve the Vlasov-Poisson system, and \cite{hahn_adaptively_2016, sousbie_coldice_2016} further developed this framework with adaptively refined phase-space tessellations. However, following the evolution of a a 3D manifold in 6D phase-space remains very computationally expensive.

A new class of neural networks use physics-based constraints to solve the data limitations and generalization problems of traditional neural networks, called Physics-Informed Neural Networks~\citep[PINNs;][]{raissi_physics-informed_2019, karniadakis_physics-informed_2021}.
Such neural networks are trained to respect any symmetry,
invariance, or conservation principles originating from the physical laws that
govern the observed data, as modeled by general time-dependent and non-linear partial differential equations. In the PINN framework, fully connected feed-forward neural networks predict target outputs in the PDE domain, using automatic differentiation~\citep{baydin_automatic_2018} to calculate the PDE derivatives analytically. This method calculates exact derivatives of the neural network on a continuous, grid-free domain.

PINNs have been sucessfully leveraged to accelerate numerical simulations in a variety of domains. They have been used to solve the Navier-Stokes equations and reproduce 3D fluid flow simulations with incredible accuracy~\citep{raissi_hidden_2020, jin_nsfnets_2021}, the Euler equations that model high-speed aerodynamic flows~\citep{mao_physics-informed_2020}, heat transfer~\citep{cai_physics-informed_2021}, and radiative transfer~\citep{mishra_physics_2021}.
PINNs have also been successfully applied to solve the Vlasov-Poisson equations in the context of Plasma simulations~\citep{zhang_physics-informed_2023}, though not for the cold cosmological case with vanishing velocity dispersion.
\par Recently, Kolmogorov-Arnold Networks \citep[KANs,][]{liu_kan_2025} have emerged as a new powerful neural architecture, alternative to traditional networks. KANs can also be used as a PDE solver (physics-informed KAN, PIKAN), with greater accuracy than regular networks on some problems \citep{shukla_comprehensive_2024}. Many applications of PIKANs focused on fluid mechanics problems, including incompressible flows \citep{jiang_solutions_2025}, compressible flows with discontinuities \citep{lei_discontinuity-aware_2025}, and flows with irregular geometries \citep{kashefi_physics-informed_2025}.

\par In this work we develop a PIKAN which predicts the displacement field $\zeta(q,\tau)$ in one dimension. The learning is fully unsupervised, and the network provides a continuous and mesh-free description of the displacement field. We first detail the network architecture and the physical constraints adopted in Section \ref{sec:methods}. We then present our results compared with a regular N-Body simulation in Section \ref{sec:results}. We discuss other applications and further developments of our work in Section \ref{sec:discussion}.

\section{Methods}\label{sec:methods}

\subsection{The singularities of CDM}

The evolution of CDM can be described by either the Eulerian or Lagrangian formulation of the PDEs in Equations~(\ref{eq:vp}) and (\ref{eq:vpL}), respectively. In \cite{kumar_physics_2023}, PINNs were developed using the Eulerian formulation of the Vlasov-Poisson equations to simulate fusion plasma with high accuracy.
Unfortunately, unlike the plasma case, the sparsity of the DM distribution $f(x,v,\tau)$ is ill-suited for a neural network representation. $f(x,v,\tau)=0$ for the vast majority of the input domain, but $f(x,v,\tau) = 1$ on a single parametric line in phase space. Neural networks struggle to learn such discontinuous and highly concentrated functions. Most strategies for modeling such instant discontinuous behaviour involve approximating the instant discontinuity with a Gaussian function~\citep{huang_solving_2021, kapoor_physics-informed_2023} or other approximations. This may be of interest for the case of warm DM, but would not accurately represent the dynamics of cold DM. We performed some initial tests to train a network to learn $f(x,v,\tau)$ using supervised learning with simulation data, but the network was unable to reach the target distribution with satisfactory accuracy. 

Therefore, we develop a PINN that solves the Lagrangian dynamics. The network predicts the 1D DM displacement field $\tilde\zeta(q,\tau)$ and is constrained by the PDE in Equation~(\ref{eq:vpL}). We use the Zel'dovich approximation to evolve the simulation from $\tau = 0$ to $\tau =0.9$, right before the first shell-crossing (the approximation being in fact exact in 1D in this regime). The network is used to predict the continued evolution of the displacement field for $\tau > 0.9$. Hence, the expression of position becomes:
\begin{equation}\label{eq:Lmap2}
x(q,\tau) =  q + \zeta_0(q) + \tilde\zeta(q,\tau)~,
\end{equation}
with $\zeta_0$ the displacement given by the Zel'dovich approximation. We use a neural network to approximate the DM displacement field given $q$ and $\tau$. We call the approximation $\tilde\zeta(q,\tau)$.

The evolution in the Lagrangian dynamics is still dominated by singularities in the density field created during each shell-crossing, resulting a discontinuous multi-stream force as shown in Figure~\ref{fig:1dnbodysim}. 

In early experiments, we used multilayer perceptrons (MLPs), a widely adopted architecture for neural networks. Each network layer $l$ has a fixed activation function $\sigma^l$, and the training only tunes the weights and biases. Each layer's output $\vec z_{l}$ is computed from the previous layer following the formula :
\begin{equation}
\vec z_{l} = \sigma^l \bigl (W^l ~ {\vec z}_{l-1} \bigr).
\end{equation}
 where $W^l$ are the weights at layer $l$. 
For activation functions, we explored using Rectified Linear Units \citep[ReLU,][]{hahnloser_digital_2000, agarap_deep_2019}, SiLU \citep{elfwing_sigmoid-weighted_2017}, ReCubeU ($x \mapsto \max(0, x^3)$), Wavelet activation \citep{herrera_wavelets_2022}, and found all these attempts to fail, as they do not allow the network to properly model the acceleration field. Cybenko's universal approximation theorem \citep{cybenko_approximation_1989} states that neural networks can model any continuous function, which is not necessarily the case for higher-order derivatives of the displacement field.

However, unlike MLPs, KANs have been proven to be able to represent not only continuous functions but also all discontinuous functions \citep{ismailov_three_2022}. We have hence devoted our attention to this class of networks.

\subsection{KAN Architecture}

KANs innovate on traditional ML frameworks by introducing learnable activation functions. Based on the Kolmogorov-Arnold representation theorem \citep{kolmogorov_representation_1957}, they use flexible univariate activation functions $\phi_{l,j,k}(\cdot)$ (that is, applied on the $k$-th component of $\vec z_{l-1}$ and contributing to the $j$-th component of $\vec z_{l}$). At each layer, $\vec z_{l}$ is now obtained through:
\begin{equation}
    \vec z_{l} = 
    \begin{pmatrix}
    \phi_{l,1,1} (\cdot)& \dots & \phi_{l,1,n_{l-1}} (\cdot)\\
    \vdots & & \vdots\\
    \phi_{l,n_{l},1} (\cdot) & \dots & \phi_{l,n_{l},n_{l-1}} (\cdot)
    \end{pmatrix}
    {\vec z}_{l-1}.
\end{equation}
The choice of a family of parametric functions to formulate the activation functions is critical for making the KAN framework viable. In the seminal work of \cite{liu_kan_2025}, each activation function (abbreviated as $\phi$) combines a SiLU with a spline:
\begin{equation}
    \phi(z) = w_1 \frac{z}{(1+e^{-z})}  + w_2 ~ \text{spline}(z),
\end{equation} with $w_1$ and $w_2$ learnable weights. 
The spline is composed of third order B-splines spread over a grid :
\begin{equation}
    \text{spline}(z) = \sum_i c_i B_i(z),
\end{equation}
where the $c_i$ are also learnable weights. 
\cite{rigas_adaptive_2024} proposed to make the gridding adaptative rather than uniform; we reuse their \textsc{JaxKAN} package that implements this feature. We choose a gridding with 16 cells, and an adaptivity parameter of $\epsilon = 0.05$.
Our input vector is $\vec z_1 = \begin{bmatrix}
   q & \tau 
\end{bmatrix}^T$, and we use 4 layers with respectively (8, 12, 8, 1) activations. This architecture has 4,536 free parameters.

\subsection{Physics-Informed Loss}

We use Lagrangian dynamics of DM to constrain the behavior of the network. Following the same procedure as in \cite{bernardeau_large-scale_2002} and \cite{rampf_cosmological_2021}, we take the Eulerian divergence of both sides of the 1D version of Equation~(\ref{eq:vpL}) to rewrite it as:
\begin{equation}
\frac{\partial}{\partial x} \biggr( \frac{\partial^2  x}{\partial \tau^2} + \frac{3}{2\tau} \frac{\partial  x}{\partial \tau}  \biggl) ~ =  - \frac{3}{2 \tau} \frac{\partial^2 \phi}{\partial x^2}
= - \frac{3}{2 \tau^2} \delta(x(q,\tau)),
\end{equation}
and then use the chain rule $\partial x/\partial q ~\nabla_x = \nabla_q$ to convert the Eulerian derivative to a Lagrangian derivative:
\begin{equation}\label{eq:starting}
\frac{\partial}{\partial q} \biggr( \frac{\partial^2  x}{\partial \tau^2} + \frac{3}{2\tau} \frac{\partial  x}{\partial \tau}  \biggl) 
~= - \frac{3}{2 \tau^2} F + \frac{3}{2 \tau^2}\biggl(\frac{\partial x}{\partial q} -1 \biggr),
\end{equation}
where $F$ is the effective multi-stream force defined as
\begin{align}
  F &=  \frac{\partial x}{\partial q} \int \hspace{-0.5em} \text{d}q' \delta_\text{D}[x(q, \tau) - x(q', \tau)] - 1 \\
   &=  \int \hspace{-0.5em} \text{d}q' \frac{\partial}{\partial q}\Theta[x(q, \tau) - x(q', \tau)] - 1,
\end{align}
and $\Theta$ is the Heaviside step function. We can write Equation~(\ref{eq:starting}) concisely in terms of the displacement $\zeta$:
\begin{equation}
\frac{\partial}{\partial q} \biggr( \frac{\partial^2  \zeta}{\partial \tau^2} + \frac{3}{2\tau} \frac{\partial  \zeta}{\partial \tau} 
-  \frac{3}{2 \tau^2} \zeta
\biggl) 
~= - \frac{3}{2 \tau^2} F,
\end{equation}
then multiply both sides by $2\tau^2/3$ and integrate with respect to $q$:
\begin{equation}\label{eq:ending}
 \frac{2 \tau^2}{3} \frac{\partial^2  \zeta}{\partial \tau^2} + \tau \frac{\partial  \zeta}{\partial \tau} - \zeta
~= - \int \hspace{-0.5em} \text{d}q' \Theta[x(q, \tau) - x(q', \tau)] - q + q_\text{I},
\end{equation}
where $q_\text{I}$ is the lower value of the domain boundary.
This serves as our PDE constraint. The derivatives on the left-hand side of Equation~(\ref{eq:ending}) can be calculated exactly using automatic differentiation.

There are several strategies for training PINNs with integro-differential equations such as Equation~(\ref{eq:ending}). For example, one can use approximate numerical techniques to solve the integral~\citep{pang_fpinns_2019}. One can also avoid integral discretization by defining auxiliary variables for integrands~\citep{yuan_-pinn_2022}, though this  introduces additional complexity to the network.
In certain cases one can define a network which solves for the antiderivative of the target distribution~\citep{lindell_autoint_2021,kumar_integral_2022}, allowing the entire PDE to be expressed in terms of higher-order derivatives. This last technique was used to solve the Vlasov-Poisson equations for plasma dynamics in~\cite{kumar_physics_2023} by solving for the antiderivative of the DM distribution $f(x,v,\tau)$ of Equation~(\ref{eq:vp}).
The same is not true of the integral on the right-hand side of Equation~(\ref{eq:ending}), which depends on the co-location of DM fluid elements and must be calculated with approximate numerical techniques.

We calculate the density on a fixed resolution grid, a strategy used by 
both N-body codes and Vlasov-Poisson solvers such as~\cite{sousbie_coldice_2016}. 
We define a grid of $N_j$ evenly sampled grid points in Lagrangian coordinates $q_j$. This sampling introduces error on the order of $L/N_j$, where $L$ is the spatial size of the domain.

With the output of the network at each of these grid points $\zeta_j \equiv \tilde \zeta(q_j,\tau)$ we define the loss as:
\begin{equation}\label{eq:PDE1}
 \mathcal{L}_\text{PDE1} = \frac{1}{N_j} \sum_{j=1}^{N_j}\Biggl( \frac{2 \tau^2}{3} \frac{\partial^2 \zeta_j}{\partial \tau^2} + \tau \frac{\partial \zeta_j}{\partial \tau} - \zeta_j + q_j - q_\text{I} + \frac{1}{N_j}\sum_{j' = 1}^{N_j} M_{j,j'} \Biggr)^2,
\end{equation}
where $M \in \mathcal{B}^{N_j \times N_j}$ is defined as
\begin{equation}
    M_{j,j'} = \left\{ 
  \begin{array}{ c l }
    1 & \quad \textrm{if } \zeta_j \geq \zeta_j' \\
    0                 & \quad \textrm{otherwise}
  \end{array}
\right.
\end{equation}
and is easily constructed through array broadcasting. 

Unfortunately the loss in Equation~(\ref{eq:PDE1}) has a large memory footprint of $\mathcal{O}(N_j^2)$ driven by the size of $M$. Therefore we also consider an alternate strategy by directly evaluating $\partial \phi / \partial x$ in Eulerian coordinates and remapping to Lagrangian coordinates.
 We first define a Eulerian grid such that $N_i = C N_j$, where $C$ is a large positive integer. We can then map each $x_j \equiv \zeta(q_j, \tau) + \zeta_0 + q_j $ to a corresponding index $i$ and position $x_i$ with an error of at most $L/N_i = L/(C N_j)$. This allows us to calculate the density directly in Eulerian space:
 \begin{equation}\label{eq:density}
\delta(x_i,\tau) = \sum_j \delta_\text{D} [x_i - x_j(\tau)] - 1,
\end{equation}

The computation of the above expression has worst-case computational complexity of $\mathcal{O}( N_i) =\mathcal{O}(CN_j)$ ~\citep{sanders_sequential_2019}. The Eulerian gradient of $\phi$ can then be calculated with a numerical integral:
\begin{equation}\label{eq:integralpde2}
\frac{\partial  \phi}{\partial x}(x_i,\tau) = \sum_{x \in x_i}^{x_i} \delta(x,\tau),
\end{equation}
 and mapped back to the Lagrangian grid by selecting the $N_j$ indexes of $i$ map back to $j$:
\begin{equation}
\frac{\partial  \phi}{\partial x}(q_j,\tau) \subseteq \frac{\partial  \phi}{\partial x}(x_i,\tau).
\end{equation}
This allows us to write Equation~(\ref{eq:vpL}) in Lagrangian coordinates despite the Eulerian derivative. The alternate PDE loss can be written as:
\begin{equation}\label{eq:PDE2}
 \mathcal{L}_\text{PDE2} = \frac{1}{N_j} \sum_{j=1}^{N_j}\Biggl( \frac{2 \tau^2}{3} \frac{\partial^2 \zeta_j}{\partial \tau^2} + \tau \frac{\partial \zeta_j}{\partial \tau} + \frac{\partial_\phi}{\partial x}(q_j,\tau) \Biggr)^2.
\end{equation}
The memory footprint of Equation~(\ref{eq:PDE2}) is $\mathcal{O}(C N_j)$, driven by the Eulerian sampling size. If $C$ is chosen such that $C < N_j$, this second PDE loss will have a smaller memory footprint compared to Equation~(\ref{eq:PDE1}). However, the interpolation to Eulerian coordinates does introduce additional error.

Note that there is no obligation to choose the same $N_j$ or $N_i$ for each timestep or even each training epoch. 

\subsection{Initial \& Boundary Conditions}

\begin{figure}
	\centering 
	\includegraphics[width=0.49\textwidth]{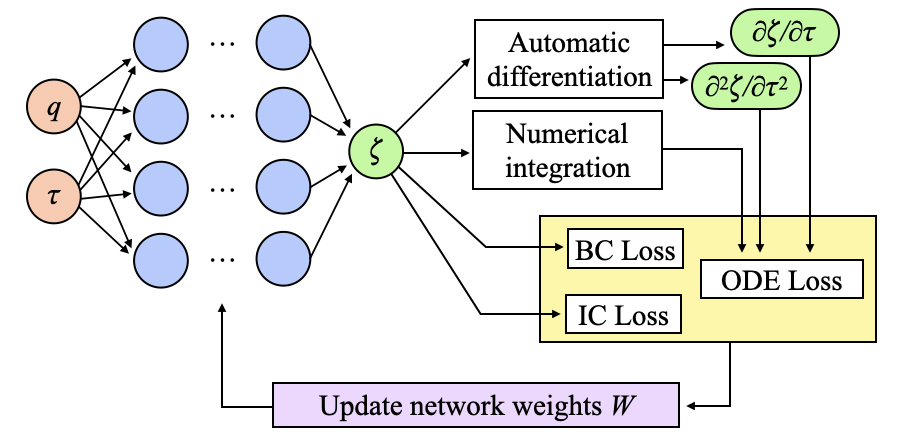}	
	\caption{Simplified schematic of the PIKAN. The input layer is represented with red nodes, hidden layers are represented with blue nodes, and the network output is represented in green. The yellow rectangles show the different components of the loss function, that drives the network optimisation (in purple).} 
	\label{fig:global_scheme}%
\end{figure}

We also impose loss terms using the initial  and boundary conditions. If the true displacement $\zeta$, velocity $\frac{\partial\zeta}{\partial\tau}$ and acceleration $\frac{\partial^2\zeta}{\partial\tau^2}$
are known at the initial time $\tau = \tau_0$, we evaluate the loss across $N_\text{IC}$ randomly chosen points in $q$:
\begin{align}\label{eq:PDEIC}
 \mathcal{L}_\text{IC} = & \frac{1}{N_\text{IC}} \sum_{j=1}^{N_\text{IC}}\Bigl( \tilde \zeta (q_j,\tau_0)-\zeta (q_j,\tau_0) \Bigr)^2 \\
 + & \frac{1}{N_\text{IC}} \sum_{j=1}^{N_\text{IC}}\Bigl( \frac{\partial\tilde \zeta (q_j,\tau_0)}{\partial\tau} - \frac{\partial\zeta (q_j,\tau_0)}{\partial\tau} \Bigr)^2 \\
 + & \frac{1}{N_\text{IC}} \sum_{j=1}^{N_\text{IC}}\Bigl( \frac{\partial^2\tilde \zeta (q_j,\tau_0)}{\partial\tau^2} - \frac{\partial^2\zeta (q_j,\tau_0)}{\partial\tau^2} \Bigr)^2.
\end{align}
Similarly, assuming that the true displacement is known at all times at the domain boundaries $q_\text{I},q_\text{F}$ we define the loss across $N_\text{BC}$ randomly chosen points in $\tau$:

\begin{align}\label{eq:BCloss}
 \mathcal{L}_\text{BC} =& \frac{1}{N_\text{BC}} \sum_{j=1}^{N_\text{BC}}\Bigl( \tilde \zeta (q_\text{F},\tau_j)-\zeta (q_\text{F},\tau_j) \Bigr)^2 \nonumber \\
 +&\frac{1}{N_\text{BC}} \sum_{j=1}^{N_\text{BC}}\Bigl( \tilde \zeta (q_\text{I},\tau_j)-\zeta (q_\text{I},\tau_j) \Bigr)^2 \nonumber \\ 
  +& \frac{1}{N_\text{BC}} \sum_{j=1}^{N_\text{BC}}\Bigl( \frac{\partial\tilde \zeta (q_\text{F},\tau_j)}{\partial \tau}-\frac{\partial \zeta (q_\text{F},\tau_j)}{\partial \tau} \Bigr)^2 \nonumber \\
 +&\frac{1}{N_\text{BC}} \sum_{j=1}^{N_\text{BC}}\Bigl( \frac{\partial \tilde \zeta (q_\text{I},\tau_j)}{\partial \tau}-\frac{\partial \zeta (q_\text{I},\tau_j)}{\partial \tau}\Bigr)^2 \nonumber \\
  +& \frac{1}{N_\text{BC}} \sum_{j=1}^{N_\text{BC}}\Bigl( \frac{\partial^2\tilde \zeta (q_\text{F},\tau_j)}{\partial \tau^2}-\frac{\partial^2 \zeta (q_\text{F},\tau_j)}{\partial \tau^2} \Bigr)^2 \nonumber \\
 +&\frac{1}{N_\text{BC}} \sum_{j=1}^{N_\text{BC}}\Bigl( \frac{\partial^2 \tilde \zeta (q_\text{I},\tau_j)}{\partial \tau^2}-\frac{\partial^2 \zeta (q_\text{I},\tau_j)}{\partial \tau^2}\Bigr)^2. 
\end{align}
We set boundary values (displacement, velocity and acceleration) to zero on both edges of the simulation domain. We summarize our PINN scheme in figure \ref{fig:global_scheme}.
\par Here we have discretised the outputs of the model only for the convenience of computing the different loss terms. However, it is important to keep in mind that the PIKAN outputs continuous fields which can then be sampled at any desired resolution, unlike N-Body techniques. This is an important advantage of our technique.

\subsection{Propagation in time}
\par Unlike usual problems tackled by PINNs, we cannot randomly and independently sample time and space coordinates in the simulation domain for computing the PDE loss term. This is because our numerical integral in equations \ref{eq:PDE1} and \ref{eq:integralpde2} require finely sampling the spatial domain a given timestep. As a result, due to memory constraints, in one batch of input points, we sample 8192 spatial coordinates and only $\sim$200 time coordinates. Even with longer training time, we found it very difficult to train a model across a long time domain, which may be due to a combination of the sparser time sampling and increased number of shell-crossings. To overcome this problem, we train an ensemble of networks sequentially on limited $\tau$ intervals. This acts like a simplified implementation of finite-basis PINNs \citep{moseley_finite_2023}. The $k$-th network starts where the $(k-1)$-th network stops, so $\tau_{k}^{start} = \tau_{k-1}^{end}$. The first network $\tilde \zeta_0$ is trained within the range $[0.9, 3]$, and each successive network $\tilde \zeta_k$ is trained on $[\tau_{k}^{start},  \tau_{k}^{start}+2]$. Initial conditions are propagated using the previous KAN solutions: 
\begin{equation}
    \zeta_k(q, \tau^{start}_{k}) = \sum_{j=1}^k \tilde \zeta_j(q, \tau^{start}_{j}), 
\end{equation}
while BCs remain unchanged. In practice, we train $k=10$ models, reaching $\tau=21$ and 6 shell-crossings.

\section{Results}\label{sec:results}

Here we present the results of using the PIKAN to predict the evolution of CDM in 1+1D phase space. We note that this method is completely unsupervised learning, using only the initial and boundary conditions with the PDE to predict the displacement field.

As a reference solution, we use the 1D N-body simulation code from \cite{rampf_unveiling_2021}. This N-body code uses a classical leapfrog integrator with a drift-kick-drift scheme at each step, without smoothening of the gravitational interactions. We run it using 8192 particles and we use the Zel'dovich approximation before $\tau=0.9$. This way, the assumptions taken for the N-Body run are similar to the conditions of the PIKAN training.

In our experiments the $\mathcal{L}_{PDE1}$ and $\mathcal{L}_{PDE2}$ losses achieved similar accuracy, though the second case trains slightly faster. Hence, in this section, we only discuss the PIKAN results obtained with the $\mathcal{L}_{PDE2}$. We show results for $\mathcal{L}_{PDE1}$ in Appendix \ref{app:resultsPDE1}.

\subsection{Phase-space and output fields}

\begin{figure*}
    \centering
    \includegraphics[trim=0cm 0cm 0cm 0cm, width=0.7\linewidth]{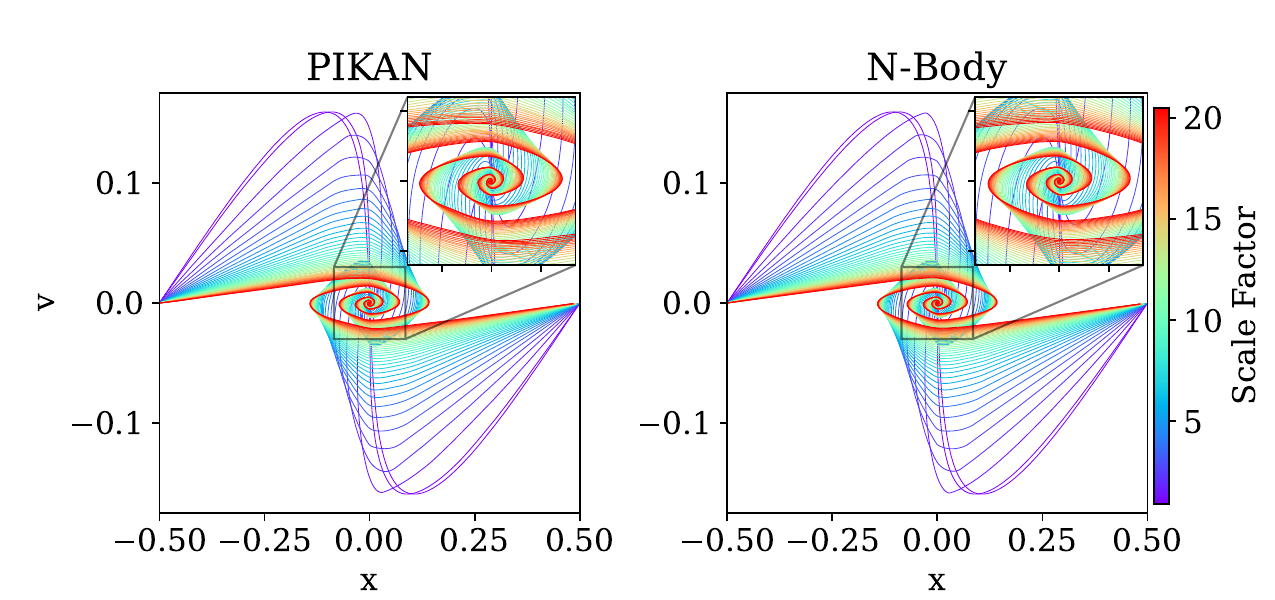}
    \caption{Phase-space distribution from the PIKAN (left) and the N-body (right) methods. Line colour indicate the conformal time from $\tau=0.9$ (purple) to $\tau=21$ (red). No data from the N-body simulation was used during the PIKAN training. The PIKAN is trained here with the $\mathcal{L}_{PDE2}$ term.}
    \label{fig:phasespacePINNvsRampf}
\end{figure*}

\begin{figure*}
    \centering
    \includegraphics[trim=0cm 0cm 0cm 0cm, width=0.9\linewidth]{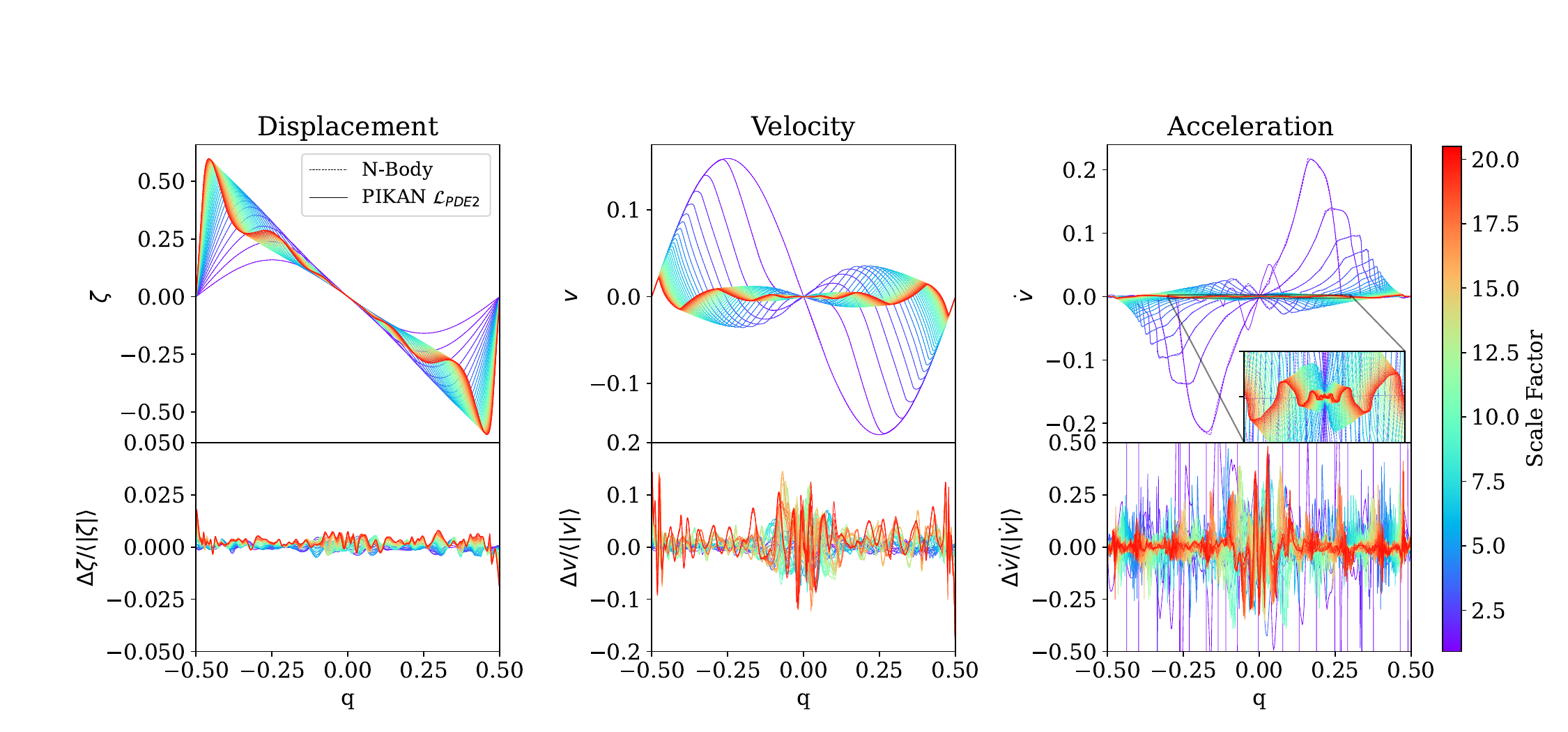}
    \caption{Output fields from the PIKAN model. The raw output is the displacement (left), and automatic differentiation provides the velocity (middle) and acceleration (right). We also show the fractional residuals with respect to the N-Body simulation (lower panels). Line colour indicates the conformal time from $\tau=0.9$ (purple) to $\tau=13$ (red). No data from the N-body simulation was used during the PIKAN training. The PIKAN is trained here with the $\mathcal{L}_{PDE2}$ term.}
    \label{fig:residualsPINNvsRampf}
\end{figure*}

\par In figure \ref{fig:phasespacePINNvsRampf} we compare the PIKAN (left) and N-body simulation (right) predictions for 2D phase-space. The velocity and acceleration for the PIKAN are obtained from automatic differentiation, while for the N-body it is provided by the code. Overall, the agreement between the PIKAN and the N-Body simulation is excellent. In figure \ref{fig:residualsPINNvsRampf} we show the displacement, velocity and acceleration fields as a function of Lagrangian coordinate $q$. The top panels show the N-Body (plain lines) and PIKAN solutions (dashed lines); by eye, they almost perfectly overlap at all times. In particular, we can see that the PIKAN is expressive enough to properly capture sharp edges and small variations in the acceleration field (see the inset panel). We also show the residuals in the bottom panels. The 95th percentile of the absolute displacement (resp. velocity) residuals increase from $0.1$\% (resp. $1.3$\%) at $\tau\approx1$ to $0.7$\% (resp. $5.6$\%) at $\tau\approx 20$. We note that at late times the greatest residuals lie on the edges ($|q|>0.4$), peaking at $1.7$\% (resp. $13$\%) for the displacement (resp. velocity), and in the center ($|q|<0.1$), reaching $0.8$\% (resp. $8$\%) for the displacement (resp. velocity). For the acceleration residuals, the 95th percentile starts at $\sim50$\% at $\tau\approx1$, before stabilizing around $\sim10$\% in the remaining of the simulation domain. For all fields, we see that the late time solutions degrades in the center of the simulation, were many shell-crossings occurred. However, it is important to note that the N-Body simulation also suffers from increased discretisation errors in the center.

\subsection{Halo profile}

\par
\begin{figure}
    \centering
    \includegraphics[width=0.9\linewidth]{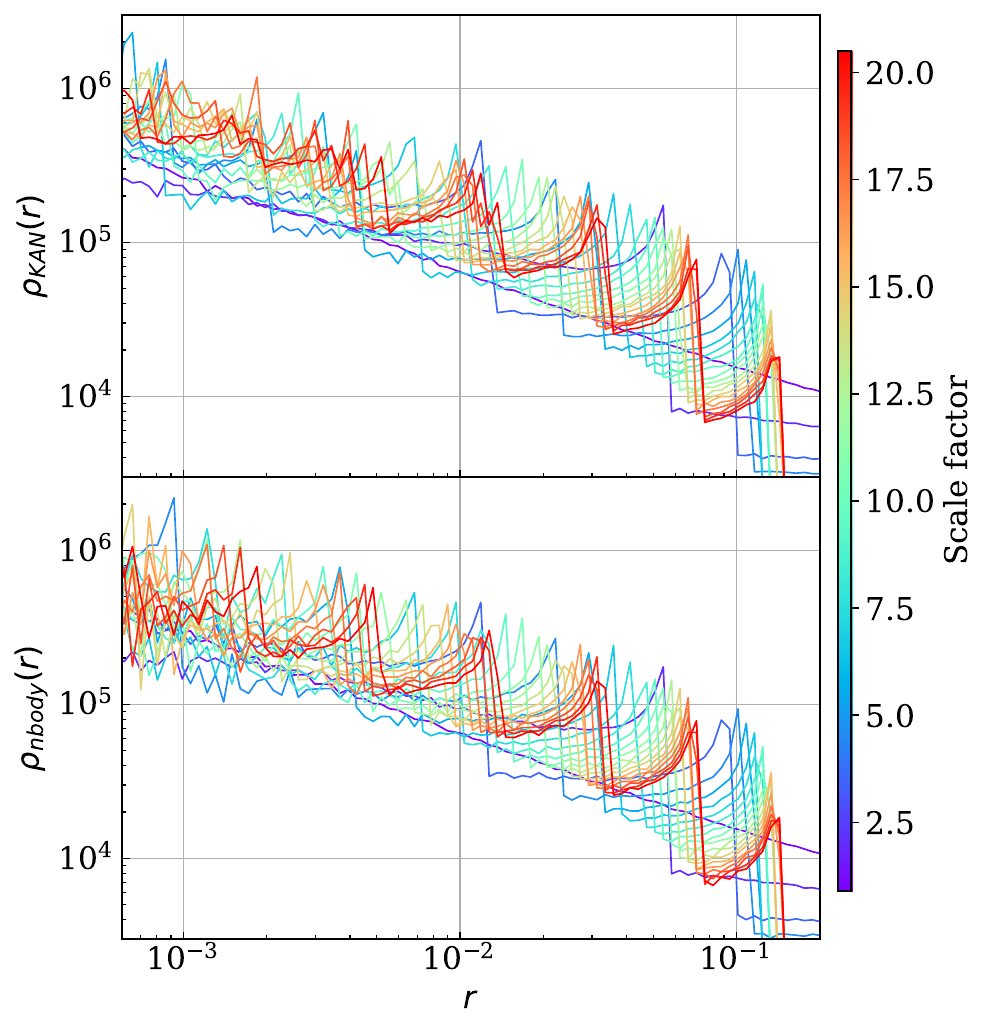}
    \caption{Comparison of the density profiles obtained through the PIKAN optimisation (top panel) and the NBody simulation (bottom panel), as a function of the cosmic time (color). The PIKAN is trained here with the $\mathcal{L}_{PDE2}$ term.}
    \label{fig:rhoprofile}
\end{figure}

 In figure \ref{fig:rhoprofile} we show the density profiles for the PIKAN (top) and N-body (bottom) simulations. We computed the profile using 8192 particles for the N-body and 8192 samples of $\zeta(q,\tau)$ for the PIKAN. The two methods are visually in good agreement. The seesaw structures reveals the several shell crossings that propagate towards the outer part of the halo. Around $r\sim10^{-3}$, the N-body profile appears very noisy, which is a natural consequence of the discretisation. The PIKAN model however seems to produce a less noisy inner profile. This is likely due to the chosen architecture, with 4 layers and 16 grids elements, that enforces some smoothness in the emulated displacement field. However, this also limits its ability to resolve well the shell crossing fronts (see for instance the upper panel around $r\sim10^{-2}$). We also stress that, once trained, the PIKAN solution can be easily upsampled to even further reduce the discreteness noise, which is not possible for a N-Body simulation run.

\subsection{Error growth}

\begin{figure}
    \centering
    \includegraphics[width=1.0\linewidth]{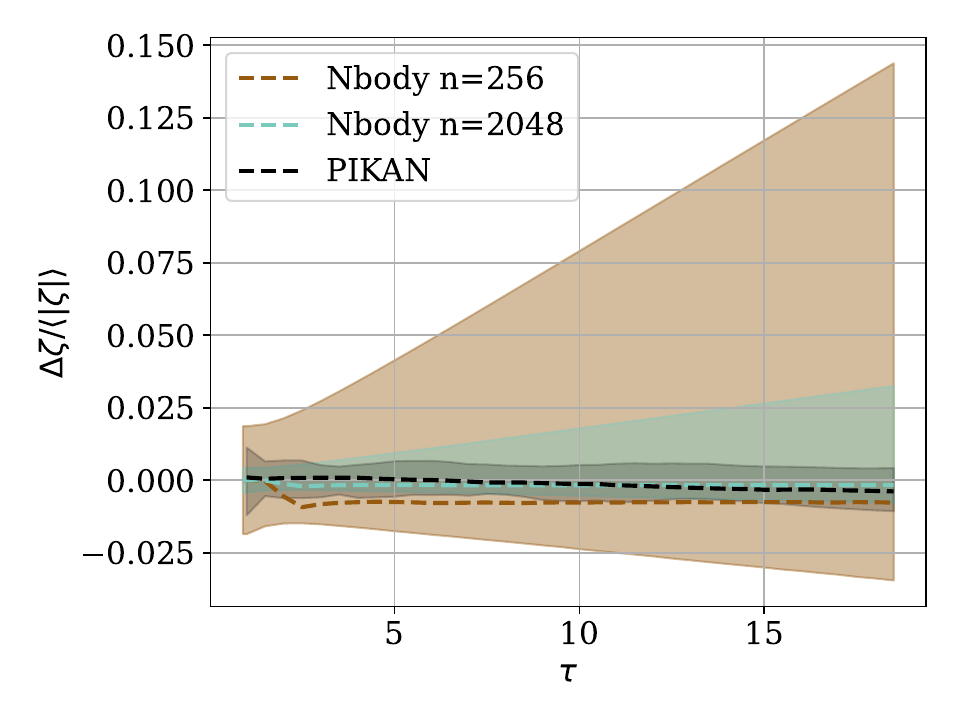}
    \caption{Fractional residuals over $\tau$ for coarse N-Body runs with 256 (brown) and 2048 (purple) particles, and for the PIKAN solution (dark). Dashed lines indicate the median and shaded regions cover 95\% of the residuals at each step.}
    \label{fig:errgrowth}
\end{figure}

Finally, we study how PIKAN and N-Body errors grows with $\tau$. We ran coarse N-Body simulations with 256 and 2048 particles, and computed the displacement fractional residuals with respect to the fiducial run (8192 particles). We show these in Figure \ref{fig:errgrowth}, in red and orange respectively for 256 and 2048 particles, along with the residuals of the PIKAN simulation, in black. Dashed lines show the median of the residuals distribution at each $\tau$ and the shaded regions encompass 95\% of the samples. We observe that the coarse N-Body runs diverge from the fiducial one. This is not a surprise as the N-Body scheme only forward propagates the PDE, accumulating discretisation errors over time. Interestingly, the spread of the PIKAN residuals remain approximately constant in the simulated time interval. This could be because the PDE optimisation is performed globally over subintervals of $\tau$, enforcing stability in the level of errors. Another possible explanation is the fact that the PIKAN learns a continuous field: CDM elements may also be locally constrained by the neighbouring flow, unlike individual particles in N-Body simulations. We highlight that this error stability is a very interesting feature of CDM PIKAN simulations, and further studies are needed to fully understand and assess this property.

\section{Discussion}\label{sec:discussion}

\begin{figure}
    \centering
    \begin{subfigure}{0.5\textwidth}
      \centering
      \includegraphics[width=0.8\textwidth]{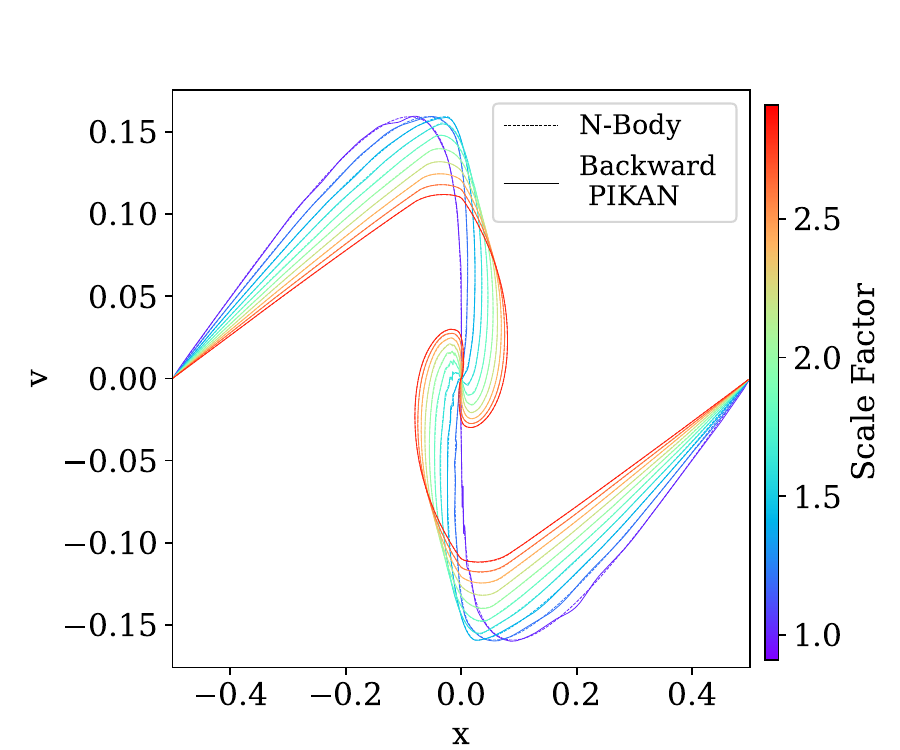}
      \caption{Backwards evolution constrained on $\zeta, \frac{\partial \zeta}{\partial\tau}$ and $\frac{\partial^2 \zeta}{\partial\tau^2}$}
      \label{fig:PINN_reversed_zvacc}
    \end{subfigure} \\
    \begin{subfigure}{0.5\textwidth}
      \centering
      \includegraphics[width=0.8\textwidth]{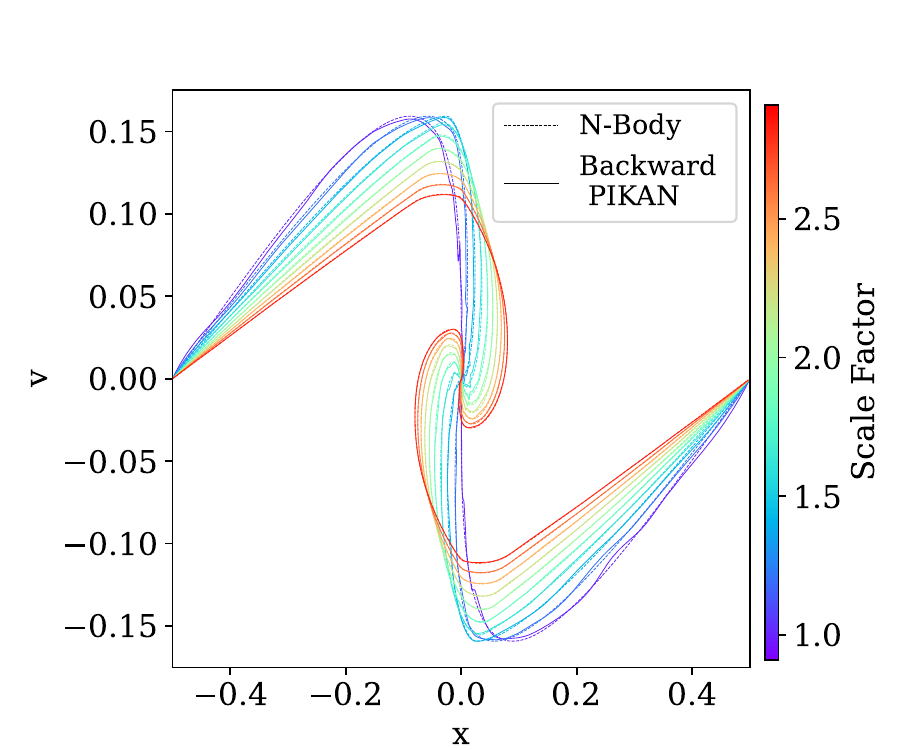}
      \caption{Backwards evolution constrained on $\zeta, \frac{\partial \zeta}{\partial\tau}$ and $\nabla  \phi$}
      \label{fig:PINN_reversed_zvdens}
    \end{subfigure}
    \caption{Phase-space distribution of the backwards-evolved PIKAN (plain lines) and the N-body (dashed lines) methods. Line colour indicates the conformal time from $\tau=0.9$ (purple) to $\tau=3$ (red). The final state of the N-body simulation serves as IC for the PIKAN training. The PIKAN is trained here with the $\mathcal{L}_{PDE2}$ term.}
    \label{fig:PINN_reversed}
\end{figure}

\subsection{Backward simulation}
The initial conditions that gave rise to the large-scale structure in the observed universe are unknown. Inferring them is an inverse problem that can be addressed with Bayesian inference \citep{jasche_bayesian_2013, legin_posterior_2023}. Here we investigate the possibility of using the PIKAN solve for these initial conditions, essentially running the simulation backwards. To implement this we can simply change the IC term of the loss function to a ``final conditions'' constraint using the displacement, velocity and acceleration fields at the final simulation time. For this test we use one KAN model and consider the time domain $\tau \in [0.9 ,  3]$. Unlike in Section \ref{sec:results} where the ICs were given by the exact Zel'dovich approximation, at $\tau = 3$ two shell-crossings have occurred and we use the N-Body solution at that time to constrain the PIKAN model.

In Figure \ref{fig:PINN_reversed_zvacc}, we show the recovered ICs in phase-space of the backwards evolution of the PIKAN, compared with the N-Body (forward) case. The evolution of the sub-manifold is overall well captured. We do however note some unrealistic wiggles in the reconstructed velocities near the initial conditions. Still, the 95th percentiles of the absolute residuals at $\tau = 0.9$ are $1.0$ \% and $4.6$ \%, respectively for displacement and velocity. The maximum error residuals respectively peak at $1.4$\% and $14$\% at early time, and are mostly located near the center of domain. It is not a surprise that the backward PIKAN optimization is a harder problem: first, the final N-Body state taken as input is approximate, and then the final (known) density field already has singularities that the model has to properly capture. Any mismodelling of the final state would propagate to previous stages of the simulation.
\par As the displacement, velocity and acceleration fields are not directly observable, it is another challenge to reconstruct the initial conditions from real large scale structure measurements. We explored the performance of the PIKAN backwards evolution under a more realistic set of late-time constraints. In figure \ref{fig:PINN_reversed_zvdens} we show the phase space distribution for when passing as final state constraint the displacement, the velocity and the gradient of the gravitational potential. $\nabla \phi$ is computed only from the observed field $x$, without requiring information on $q$. Like in the previous case, the phase space before the first shell crossing is globally well modelled. We also note the presence of spurious wiggles, which amplitude appears greater than in Fig. \ref{fig:PINN_reversed_zvacc}. Checking the residuals confirms the visual inspection: on all fields they are approximately doubled with respect to the evolution which uses the final-state acceleration. 

We also tried the backwards evolution by replacing the displacement field by the overdensity field in the input constraints, without success. We expect that the inversion of two shell crossings is a difficult task in the presence of noisy or incomplete IC. Bayesian methods may constitute a path to improve this backwards evolution, and they can be interfaced with a PINN framework \citep{yang_b-pinns_2021}.

\subsection{Current limitations and perspectives}

Scaling our method from 1D to 3D and to large cosmological simulations is the main challenge for future work. The major drawback of KAN models is that they are expensive to train. This was the principal factor limiting the complexity of our simulation case. In this study, each PIKAN model is trained for $\sim$10 hGPU, meaning that the longest run presented in Section \ref{sec:results} required around 100 hGPU. This is a problem since the optimisation relies on a given set of ICs: a given trained network cannot generalise to new initial conditions. 
Autoregressive models may provide a solution to train networks independently of a specific IC \citep[see][]{kohl_benchmarking_2024}. It may also be possible to improve the training speed with distributed training and numerical optimisation. We discuss in Appendix \ref{app:sampling} the impact of spatial sampling for optimising the PDE residuals.

\par Looking carefully at the displacement residuals in Figure \ref{fig:residualsPINNvsRampf}, one may notice a wavy pattern, meaning that the errors are correlated on a specific scale. This is due to the gridding of the spline activations in the KAN layers, and it implies the solution to be smooth below the gridding scale. In fact, as the number of shell crossings increases with $\tau$, small-scale features appear in the fields of interest (see, for instance, the zoom on the acceleration field in Figure \ref{fig:residualsPINNvsRampf}). Even with grid adaptivity, at some point the PIKAN no longer has sufficient resolution to capture the small-scale and low amplitude features caused by the last shell crossings. Adding more layers and increasing the gridding are the simplest solutions to overcome this effect, but naturally this slows down the training. It is also important to keep in mind that the KAN smoothness has the benefit of allowing upsampling of the emulated fields after training.
\par While we focused in this work on KANs built with B-splines, new basis functions have been proposed including Jacobi polynomials \citep{guo_physics-informed_2025, kashefi_pointnet_2025}, wavelets \citep{bozorgasl_wav-kan_2024} and Fourier series \citep{li_kolmogorovarnold_2025}. Future studies could try to apply these other approaches and test their ability to propagate further the CDM simulation.

\section{Summary and conclusions}

In this work we demonstrated the first application of physics-informed KANs, a novel class of neural networks with learnable activation functions, for evolving CDM fields. We considered a one-dimensional halo collapse and compared the results with a classical N-Body method. The PIKAN model is constrained solely by the Vlasov-Poisson equations, without using any N-body data during training. The PIKAN predicts the displacement fields, with velocity and acceleration obtained by automatic differentiation. To compute the loss, we solve numerically the Poisson equation and evaluate the residuals of the Vlasov equation. We derived two formulas to integrate the density, through a Lagrangian or Eulerian approach. Sequential trainings of PIKAN models allow us to propagate the solution up to 6 shell-crossings.
\par The PIKAN solution solution agrees very well with N-Body results, for the displacement, velocity and acceleration fields. In particular, the displacement error is below the percent level. We also showed that the density profiles obtained through PIKAN optimisation are also in good agreement with N-Body simulations. An important asset of our approach is that it provides a resolution-free solution. We also observed that the displacement residuals of the PIKAN remain stable, unlike N-body schemes where discretisation and forward propagation increase the error. This is also an interesting advantage of our approach.
\par Moreover, we tested the PIKAN in a backward simulation scheme. Even if less accurate than the forward training, the PIKAN is still able to inverse the effect of two shell crossings, once again only through physical contraints. Further work is needed to assess if this approach is feasible with noisier and less abundant observables, and if this can be used together with Bayesian methods to reconstruct initial conditions of a dark matter field.
\par This study constitutes a proof-of-concept for a novel way of evolving DM fields, independently from N-Body simulations. Several challenges remain to extend this method to higher dimensions and greater volumes maintaining an acceptable computational cost. Future work may focus on autoregressive approaches and numerical optimisation, with the ultimate goal of applying this framework to modelling large-scale structure or reconstructing initial conditions in upcoming cosmological surveys, independently from N-Body methods.

\section*{Acknowledgements}

NC, ET, and AM acknowledge support from the Swiss National Science Foundation under the SNSF Starting Grant ``Deep Waves'' (218396).
The authors thank Romain Teyssier and Oliver Hahn for useful discussions around this work. This work used computational resources of the Swiss National Computing Center (CSCS), through the allocated project sk030. 

\section*{Data Availability}
The base of our code and the model presented in section \ref{sec:results} are available in the \textsc{cdm-pikan}\footnote{\url{https://github.com/nicolas-cerardi/cdm-pikan}} public repository. Other models and their results will be shared upon reasonable request to the authors.



\bibliographystyle{mnras}
\bibliography{example} 




\appendix

\section{Results obtained with $\mathcal{L}_{PDE1}$}\label{app:resultsPDE1}

\begin{figure*}
    \centering
    \includegraphics[trim=0cm 0cm 0cm 0cm, width=0.9\linewidth]{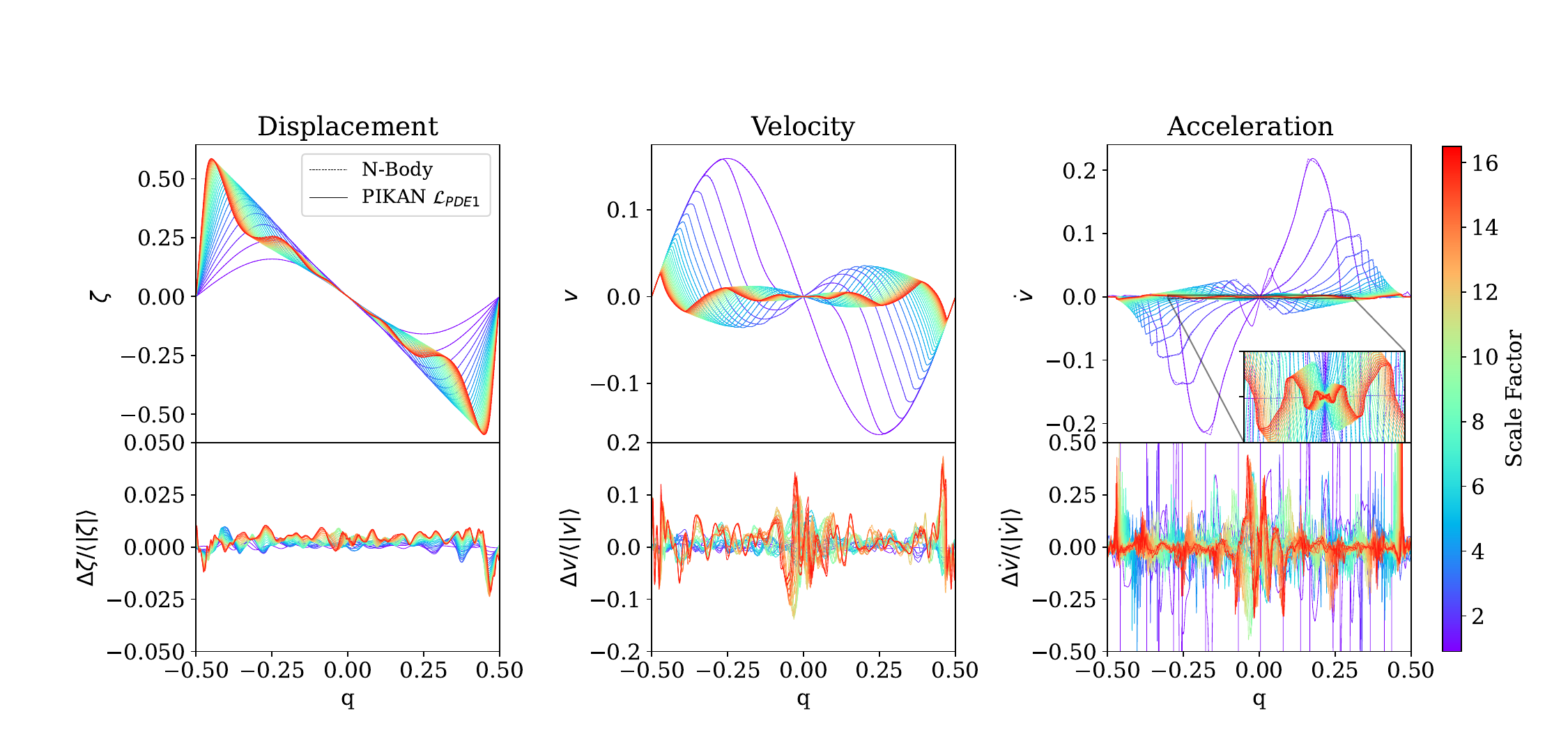}
    \caption{Output fields from the PIKAN model. The raw output is the displacement (left), and automatic differentiation provides the velocity (middle) and acceleration (right). We also show the fractional residuals with respect to the N-Body simulation (lower panels). Line colour indicates the conformal time from $\tau=0.9$ (purple) to $\tau=13$ (red). No data from the N-body simulation was used during the PIKAN training. The PIKAN is trained with the $\mathcal{L}_{PDE1}$ term.}
    \label{fig:phasespacePINNvsRampf_pde1}
\end{figure*}

We here present results from a PIKAN optimisation with the $\mathcal{L}_{PDE1}$ loss, with otherwise similar conditions than in Section \ref{sec:results}. We recall that $\mathcal{L}_{PDE1}$ is the result of a Lagrangian approach when integrating the discretised density field. This derivation is in principle more accurate but slower than its Eulerian counterpart $\mathcal{L}_{PDE2}$. In figure \ref{fig:phasespacePINNvsRampf_pde1} we compare the PIKAN (upper row) and N-body simulation (bottom row) outputs. We show the phase space (left), position (center left), velocity (center right) and acceleration (right). Infact, the $\mathcal{L}_{PDE1}$ optimisation yields similar final residuals than the $\mathcal{L}_{PDE2}$ training, for all fields. The residuals with respect to the N-Body simulation are very similar. The computational time is increased by about $10$\%. when using $\mathcal{L}_{PDE1}$. All in all, this indicates that with the good resolution adopted for the Eulerian approach (computing the density fields in $8192\times500$ cells), both losses perform equivalently.

\section{Variations of spatial sampling}\label{app:sampling}

\begin{figure}
    \centering
    \includegraphics[trim=0cm 0cm 0cm 0cm, width=0.9\linewidth]{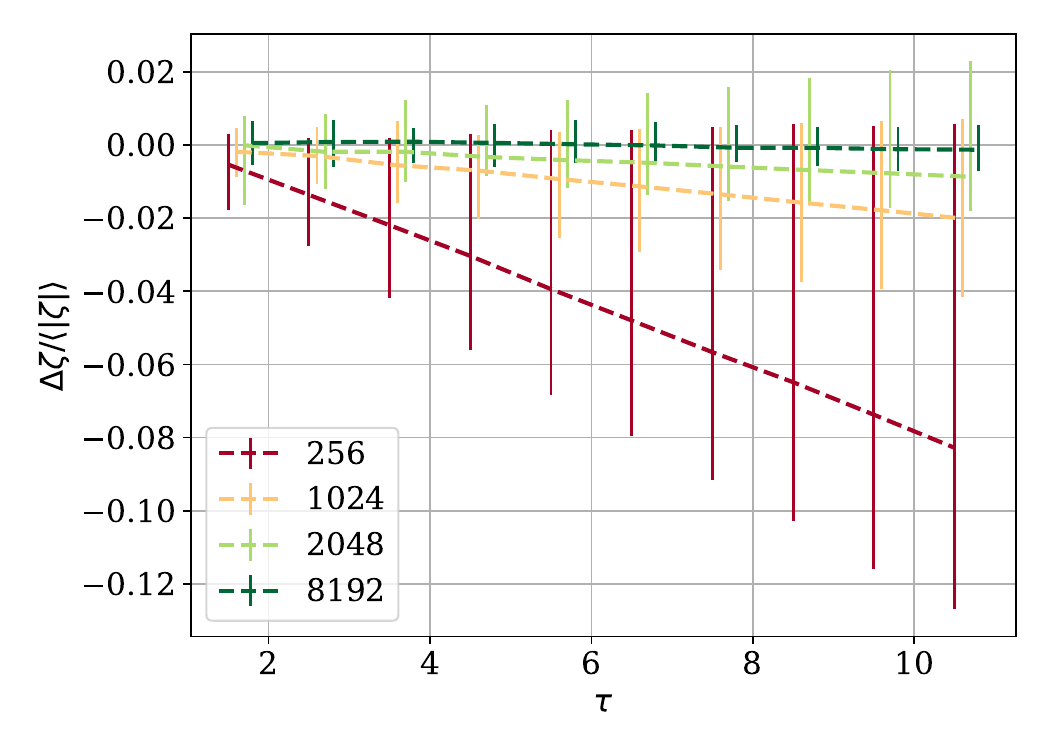}
    \caption{Residuals as a function of $\tau$ for PIKAN models with different PDE sampling size $N_j$ (see equation \ref{eq:PDE2}), increasing from $256$ (red) to $8192$ (dark green). Dashed lines indicate the median of the residuals and the error bars show the regions containing 95\% of the samples. The PIKAN is trained with the $\mathcal{L}_{PDE2}$ term.}
    \label{fig:error_vs_sampling}
\end{figure}

\begin{figure}
    \centering
    \includegraphics[trim=2cm 0cm 3cm 1cm, width=0.9\linewidth]{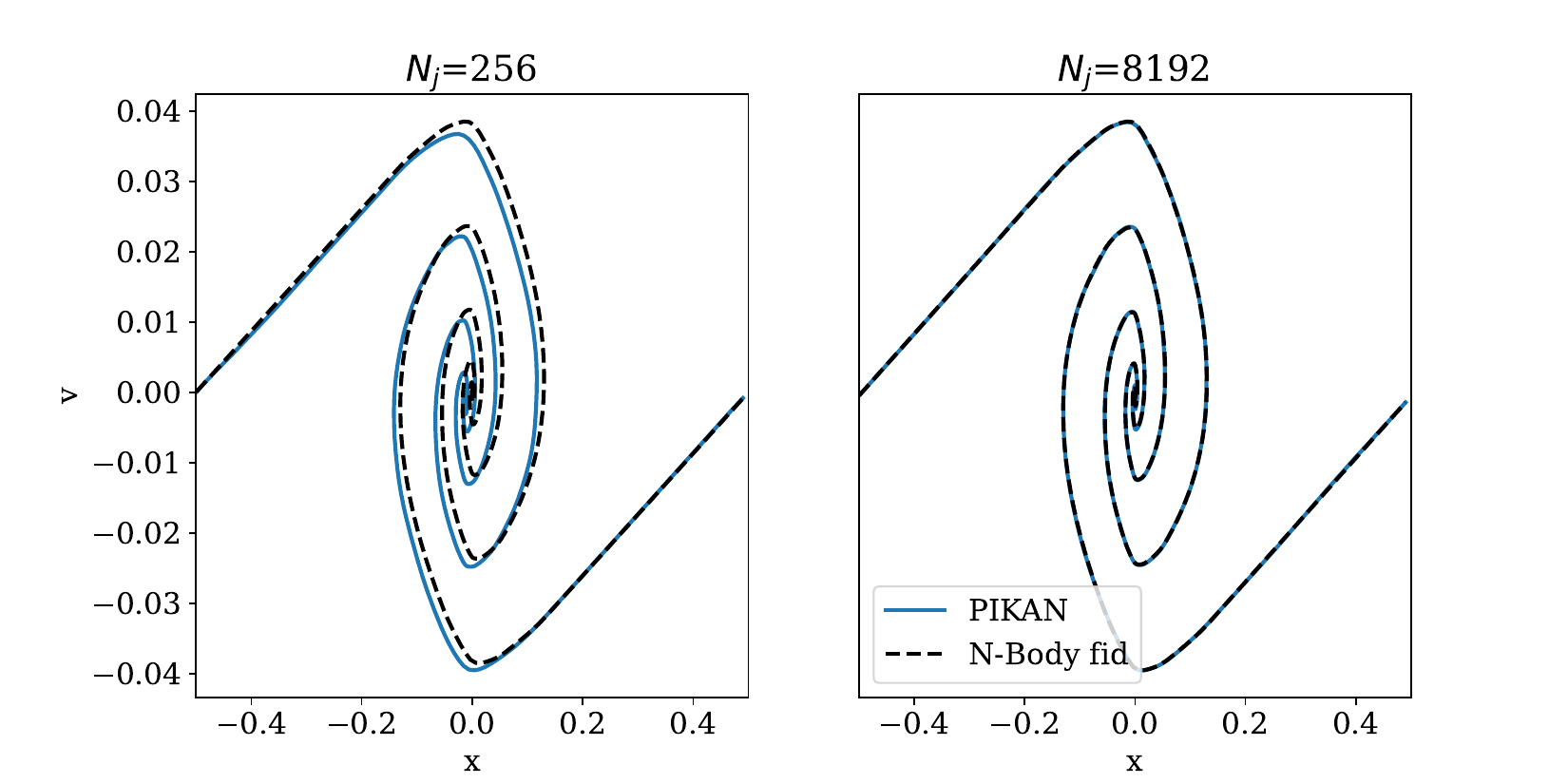}
    \caption{Phase space distribution at $\tau=18.5$ for the PIKAN solution (blue) with PDE sampling size $N_j=256$ (left) and $N_j=8192$  (right). Also shown as reference is the fiducial N-Body run with 8192 particle (dashed black).}
    \label{fig:lowres_shift}
\end{figure}

In section \ref{sec:results}, the PDE loss was computed by sampling $N_j=8192$ spatial coordinates at each evaluated timestep. This number was chosen to be the same than the number of particles in the fiducial N-Body simulation. Here, we test the effect of downsampling the PDE loss on the displacement residuals. We trained models using $\mathcal{L}_{PDE2}$ with $N_j=256, 512, 1024$ and $2048$ up to $\tau=11$, and compare with the PIKAN model from section \ref{sec:results}. The sampling is always uniform in $q$. In Figure \ref{fig:error_vs_sampling}, we show how the displacement residuals evolves with $\tau$. Plain lines indicate the median and shaded regions encompass 95\% of the residuals. Interestingly, all downsampled model have a bias growing with $\tau$. Still, the solution of the $N_j=2048$ model remains within 2\% of the fiducial N-Body run. As the training time is proportional to $N_j$, this shows that it is possible to reduce $N_j$ by a factor of 2 or 4, depending on the requirements in accuracy. The presence of a bias at low resolution is well visible in figure \ref{fig:lowres_shift}. At lower resolution, the CDM flows follows the same spiral collapse in phase space, but with a global drift towards negative $x$. We investigated the residuals as a function of $q$ and found that there are important, antisymmetric errors near the edges of the domain, which amplitude increases when $N_j$ is reduced and $\tau$ grows. This could be responsible for the measured bias. Hence, it suggests that a non-uniform sampling strategy may focus on fine sampling the domain boundaries together with the shell crossings positions. Another way to tackle this issue could be to enforce hard BC constraints, e.g. setting the PIKAN solution to 0 on the domain edges instead of optimising $\mathcal{L}_{BC}$.

\bsp	
\label{lastpage}
\end{document}